\input harvmac.tex

\input epsf.tex

% \draftmode

\def\figin{\epsfcheck\figin}\def\figins{\epsfcheck\figins}
\def\epsfcheck{\ifx\epsfbox\UnDeFiNeD
\message{(NO epsf.tex, FIGURES WILL BE IGNORED)}
\gdef\figin##1{\vskip2in}\gdef\figins##1{\hskip.5in}% blank space instead
\else\message{(FIGURES WILL BE INCLUDED)}%
\gdef\figin##1{##1}\gdef\figins##1{##1}\fi}
\def\DefWarn#1{}
\def\figinsert{\goodbreak\midinsert}
\def\ifig#1#2#3{\DefWarn#1\xdef#1{fig.~\the\figno}
\writedef{#1\leftbracket fig.\noexpand~\the\figno}%
\figinsert\figin{\centerline{#3}}\medskip\centerline{\vbox{\baselineskip12pt
\advance\hsize by -1truein\noindent\footnotefont{\bf
Fig.~\the\figno:} #2}}
\bigskip\endinsert\global\advance\figno by1}
%%% TO PUT FIGURES INSERT:
%\ifig\fivelegs{ Write caption
%  } {\epsfxsize2.5in\epsfbox{fivelegs.eps}}

%%%% definitions

%%%%%%%%%%%%%%%%%%%%%%%%%%%%%%%%%%%%%%%%5
%\ColemanAW
\lref\CdeL{
  S.~R.~Coleman and F.~De Luccia,
  %``Gravitational Effects On And Of Vacuum Decay,''
  Phys.\ Rev.\  D {\bf 21}, 3305 (1980).
  %%CITATION = PHRVA,D21,3305;%%
}

%\AbbottKR
\lref\AbbottKR{
  L.~F.~Abbott and S.~R.~Coleman,
  %``The Collapse Of An Anti-De Sitter Bubble,''
  Nucl.\ Phys.\  B {\bf 259}, 170 (1985).
  %%CITATION = NUPHA,B259,170;%%
}

 % Talks about dS field theories and their AdS duals.
%\BuchelWF
\lref\BuchelWF{
  A.~Buchel,
  %``Gauge / gravity correspondence in accelerating universe,''
  Phys.\ Rev.\  D {\bf 65}, 125015 (2002)
  [arXiv:hep-th/0203041].
  %%CITATION = PHRVA,D65,125015;%%
}

% AdS/CFT  on dS with mass deformation
%\BuchelKJ
\lref\BuchelKJ{
  A.~Buchel, P.~Langfelder and J.~Walcher,
  %``On time-dependent backgrounds in supergravity and string theory,''
  Phys.\ Rev.\  D {\bf 67}, 024011 (2003)
  [arXiv:hep-th/0207214].
  %%CITATION = PHRVA,D67,024011;%%
}
%\HirayamaJN
\lref\HirayamaJN{
  T.~Hirayama,
  %``A holographic dual of CFT with flavor on de Sitter space,''
  JHEP {\bf 0606}, 013 (2006)
  [arXiv:hep-th/0602258].
  %%CITATION = JHEPA,0606,013;%%
}

%\PolyakovNQ
\lref\Polyakov{
  A.~M.~Polyakov,
  %``Decay of Vacuum Energy,''
  Nucl.\ Phys.\  B {\bf 834}, 316 (2010)
  [arXiv:0912.5503 [hep-th]].
  %%CITATION = NUPHA,B834,316;%%
}
%\WittenGJ
\lref\WittenKK{
  E.~Witten,
  %``Instability Of The Kaluza-Klein Vacuum,''
  Nucl.\ Phys.\  B {\bf 195}, 481 (1982).
  %%CITATION = NUPHA,B195,481;%%
}

%\MaldacenaRE
\lref\AdSCFT{
  J.~M.~Maldacena,
  %``The large N limit of superconformal field theories and supergravity,''
  Adv.\ Theor.\ Math.\ Phys.\  {\bf 2}, 231 (1998)
  [Int.\ J.\ Theor.\ Phys.\  {\bf 38}, 1113 (1999)]
  [arXiv:hep-th/9711200].
  %%CITATION = IJTPB,38,1113;%%
 E.~Witten,
  %``Anti-de Sitter space and holography,''
  Adv.\ Theor.\ Math.\ Phys.\  {\bf 2}, 253 (1998)
  [arXiv:hep-th/9802150].
  %%CITATION = 00203,2,253;%%
 S.~S.~Gubser, I.~R.~Klebanov and A.~M.~Polyakov,
  %``Gauge theory correlators from non-critical string theory,''
  Phys.\ Lett.\  B {\bf 428}, 105 (1998)
  [arXiv:hep-th/9802109].
  %%CITATION = PHLTA,B428,105;%%
}

%\AlishahihaMD
\lref\AlishahihaMD{
  M.~Alishahiha, A.~Karch, E.~Silverstein and D.~Tong,
  %``The dS/dS correspondence,''
  AIP Conf.\ Proc.\  {\bf 743}, 393 (2005)
  [arXiv:hep-th/0407125].
  %%CITATION = APCPC,743,393;%%
}

%% AdS/CFT on de-Sitter .

%\HawkingDA
\lref\HawkingDA{
  S.~Hawking, J.~M.~Maldacena and A.~Strominger,
  %``DeSitter entropy, quantum entanglement and AdS/CFT,''
  JHEP {\bf 0105}, 001 (2001)
  [arXiv:hep-th/0002145].
  %%CITATION = JHEPA,0105,001;%%
}

%\BuchelIU
\lref\BuchelIU{
  A.~Buchel and A.~A.~Tseytlin,
  %``Curved space resolution of singularity of fractional D3-branes on
  %conifold,''
  Phys.\ Rev.\  D {\bf 65}, 085019 (2002)
  [arXiv:hep-th/0111017].
  %%CITATION = PHRVA,D65,085019;%%
}

%\BuchelWF
\lref\BuchelWF{
  A.~Buchel,
  %``Gauge / gravity correspondence in accelerating universe,''
  Phys.\ Rev.\  D {\bf 65}, 125015 (2002)
  [arXiv:hep-th/0203041].
  %%CITATION = PHRVA,D65,125015;%%
}
%\BuchelTJ
\lref\BuchelTJ{
  A.~Buchel, P.~Langfelder and J.~Walcher,
  %``Does the tachyon matter?,''
  Annals Phys.\  {\bf 302}, 78 (2002)
  [arXiv:hep-th/0207235].
  %%CITATION = APNYA,302,78;%%
}
%\HirayamaJN
\lref\HirayamaJN{
  T.~Hirayama,
  %``A holographic dual of CFT with flavor on de Sitter space,''
  JHEP {\bf 0606}, 013 (2006)
  [arXiv:hep-th/0602258].
  %%CITATION = JHEPA,0606,013;%%
}

%\BuchelEM
\lref\BuchelEM{
  A.~Buchel,
  %``Inflation on the resolved warped deformed conifold,''
  Phys.\ Rev.\  D {\bf 74}, 046009 (2006)
  [arXiv:hep-th/0601013].
  %%CITATION = PHRVA,D74,046009;%%
}

%%%%%%%%%%%%%%%%%%%%%%%%%%%%%%%%%%%%%%%%%%%%%%%%%%%%%%%%%%%%%%%%%%%%
\Title{\vbox{\baselineskip12pt \hbox{} \hbox{
} }} {\vbox{\centerline{ Vacuum decay into Anti de Sitter space
  }
\centerline{
  }
}}
\bigskip
\centerline{  Juan Maldacena }
\bigskip
\centerline{ \it  School of Natural Sciences, Institute for
Advanced Study} \centerline{\it Princeton, NJ 08540, USA}

\vskip .3in \noindent
%%%%%%%%%%%%%%%%%%%%%%%%%%%%%%%%%%%%%%%%%%%%%%%%%%%%%%%%%%%%%%%%%%%%%%%%%%%%%%%%%%%%%%%%%%%%

We propose an interpretation of decays of a false vacuum into an $AdS$ region.
The $AdS$ region is interpreted in terms of a dual field theory living on an end of
the world brane which expands into the false vacuum.

%%%%%%%%%%%%%%%%%%%%%%%%%%%%%%%%%%%%%%%%%%%%%%%%%%%%%%%%%%%%%%%%%%%%%%%%%%%%%%%%%%%%%%%%%%%%

 \Date{ }

%%%%%%%%%%%%%%%%%%%%%%%%%%%%%%%%%%%%%%%%%%%%%%%%%%%%%%%%%%%%%%%%%%%%%%%%%

\newsec{ Introduction}

In a  classic paper \CdeL ,  Coleman and de Luccia  studied vacuum decay in
theories of gravity. They pointed out that decays into $AdS$ generically produce
a singularity. This has been interpreted by some authors as an obstacle for applying
AdS/CFT \AdSCFT\ to this problem. Here we point out that AdS/CFT can be applied
to this problem in a way that   gives a non-singular description of the decay process.

The idea is very simple. The bubble decay geometry contains only a portion of $AdS$.
This portion can be interpreted as a field theory with a cutoff. The field theory lives on
the domain wall. The domain wall has a de-Sitter geometry. Thus we have a field theory on
de Sitter space. A field theory on de-Sitter space is well defined, provided we choose the
standard Euclidean vacuum. Thus, this field theory gives a dual description of the whole interior
of the domain wall. When we are away from the thin wall approximation, the conformal field theory
is perturbed by an irrelevant (or relevant) operator.   This picture is most clear in cases where there is a parametric
separation between the radius of $AdS$ and the radius of curvature of the de-Sitter expanding geometry.

The conclusion is that decays into $AdS$ are conceptually similar to decays via end of the world branes, or
bubbles of nothing \WittenKK .

\newsec{ A static domain wall}

\ifig\FlatCase{ In (a) we see a thin wall brane joining flat space to an $AdS$ space.
In (b) we replace the $AdS$ region by a CFT that lives on an end of the world brane.
In (c) we consider a thick wall brane. The scalar field approaches the $AdS$ minimum only asymptotically.
 In (d) that is replaced by an end of the world brane
plus a CFT.  The fact that the scalar field is away from the minimum implies that
  the CFT is perturbed by an irrelevant deformation.
  } {\epsfxsize3.0in\epsfbox{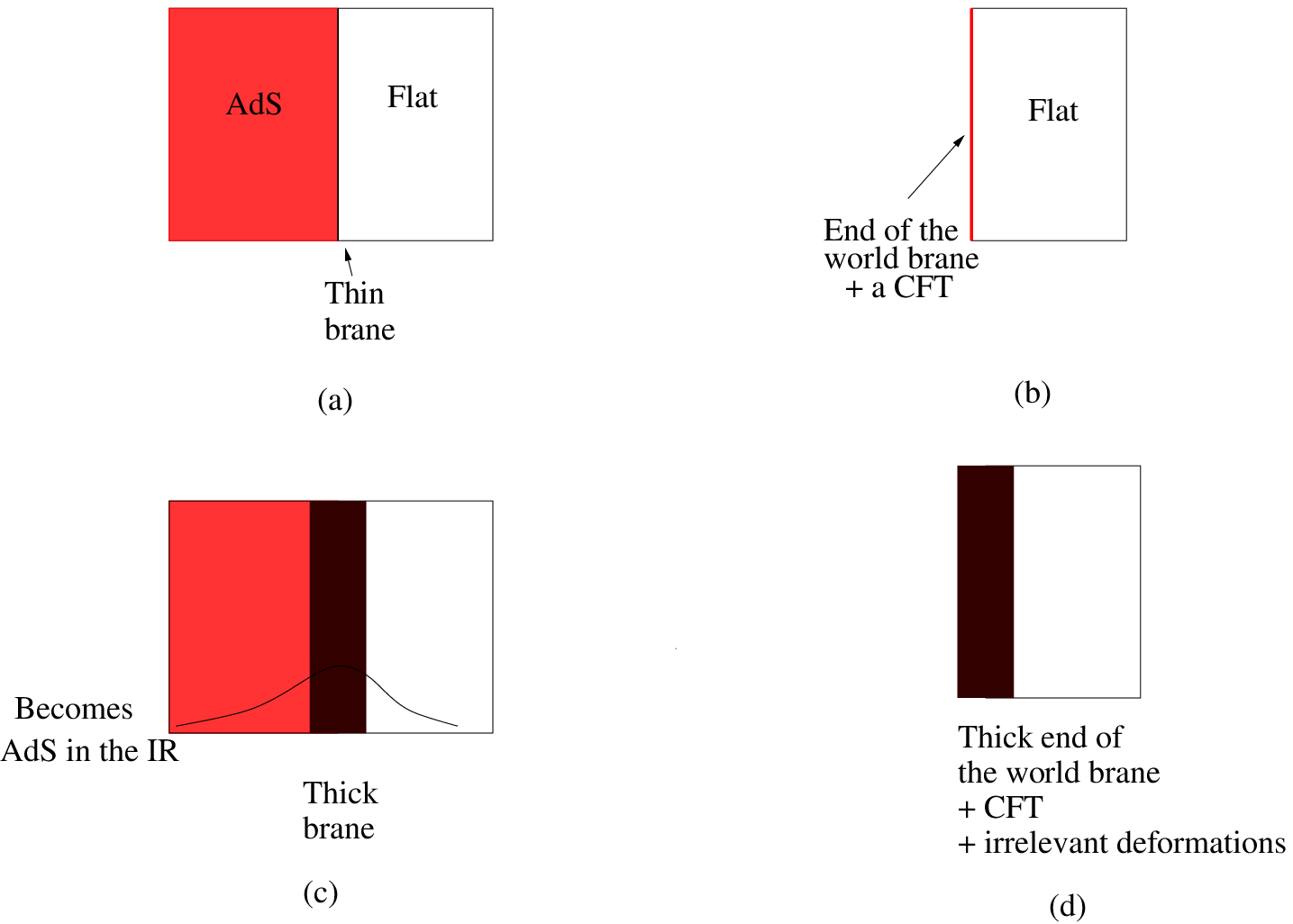}}

Here we consider a static (non-expanding)
domain wall separating an $AdS$ region from flat space.
Let us start in the thin wall approximation.
We  adjust the tension to a critical value, $T_{cr}$,
 in order order to have a static domain wall\foot{ The critical
 tension
is $ T_{cr}  = 12 { M_{pl}^2 \over R_{AdS_4} } \sim  M_{pl} \sqrt{\Lambda_{ADS} } $.
}.
 With a domain wall at $\rho=0$, we have
 Minkowski space for $\rho>0$ and $AdS$ space for $\rho<0$. The $AdS$ metric is
 \eqn\adsmetr{
 ds^2 =  d\rho^2 + e^{  2 \rho /R_{AdS} } dx_{2+1}^2 ~~,~~~~~~~{\rm for}~~\rho<0
  }
  Via the AdS/CFT duality, we can now replace the $AdS$ region by a conformal field theory. See
  \FlatCase .
  Since we only get a portion of $AdS$ space, we get a field theory with a UV cutoff. This field theory
  becomes conformal at low energies. Thus the whole set up can be viewed as flat space with
  an ``end of the world"  brane. On this end of the world brane we have a set of degrees of freedom which
  becomes conformal at low energies.   In the coordinates in
  \adsmetr\ the UV cutoff on the CFT is of the order of the radius of $AdS$, $R_{AdS}$.
 The rough number
  of degrees of freedom of the CFT is $  c \propto R^2_{AdS} M_{pl}^2 $, where
  $M_{pl}$ is the Planck mass.

  Let us now move away from the thin wall approximation and consider a potential with two minima, one
  with zero potential energy and one with a negative potential energy. The potential has to be tuned
  so that we get a static domain wall. There is  a transition region joining the flat space and the
  $AdS$ space region. The scalar field is non-zero throughout   $AdS$, but it approaches
  zero as $\rho \to -\infty$, which is the IR region. This can be interpreted as a field theory
  with a UV cutoff, which is perturbed by an irrelevant operator (dual to the scalar field) at the
  cutoff scale.
  The detailed shape of the domain wall is telling us how the field theory is coupled
  to the flat space region at the cutoff scale. (Supersymmetric domain walls of this kind were
  studied in %\CveticBF
\ref\CveticBF{
  M.~Cvetic, S.~Griffies and S.~J.~Rey,
  %``Static domain walls in N=1 supergravity,''
  Nucl.\ Phys.\  B {\bf 381}, 301 (1992)
  [arXiv:hep-th/9201007].
  %%CITATION = NUPHA,B381,301;%%
}.)

  Of course, one can also consider an end of the world brane which does not have many degrees
  of freedom in the IR, such as an orbifold plane (which has none), or a brane with
   a gauge theory with a small number of
  colors, or with a field theory which is gapped and not conformal in the IR. A string theory example
  of an end of the world brane with a CFT in the IR is the Horava-Witten end of the world brane
  %\HoravaQA
\ref\HoravaQA{
  P.~Horava and E.~Witten,
  %``Heterotic and type I string dynamics from eleven dimensions,''
  Nucl.\ Phys.\  B {\bf 460}, 506 (1996)
  [arXiv:hep-th/9510209].
  %%CITATION = NUPHA,B460,506;%%
}
in a
  $T^6$ compactification to five  dimensions. At low energies we have an $E_8$, ${\cal N}=4$ super Yang Mills
  on the brane. Here the end of the world brane is
  extended along the $T^6$ and four of the five dimensions.

 \newsec{An expanding domain wall}

 We now consider the case where the tension is very slightly below the critical value $T<T_{cr}$ \foot{
 If the tension of the domain wall is $T>T_{cr}$ the decay cannot proceed \CdeL .}.
 In this case the domain wall   expands. The geometry of the domain wall is a three dimensional
 de-Sitter space
 of radius of curvature $R_{dS}$. It is convenient to slice the space with $dS_3$ slices to obtain
 a full geometry of the form \CdeL
 \eqn\interio{\eqalign{
  ds^2 = &  R^2_{AdS} [  d\rho^2 + \sinh^2 \rho ds^2_{dS_3} ] ~,~~~~~~~~~ \rho < \rho_0
  \cr
  ds^2 = & dr^2 + r^2 ds^2_{dS_3} ~,~~~~~~~~~~~~~~~~~~~~~~~ r_0 < r
  \cr
  & R_{dS} \equiv r_0 = R_{AdS} \sinh \rho_0  ~,~~~~~~~~~~~~~~ e^{-\rho_0} = { T_{cr} - T \over T_{cr} +T}
  }}
   where the value of $\rho_0$ is obtained from the junction conditions. Here $ds^2_{dS_3}$ is just the metric
   of three dimensional de-Sitter with unit radius. And $T_{cr}$ is the critical tension for a given
   $R_{AdS}$.

   When the tension of the domain wall is very close to the critical tension  the
   expansion of the wall is very slow.
   In this regime, it is clear that we can think of the domain wall again as an end of the
   world brane plus a conformal field theory. The end of the world brane is
   expanding, because it has a non-zero effective tension, $\tilde T <0$. The effective cutoff on the
   end of the world brane is again of the order of $1/R_{AdS}$. If $\rho_0$ is very large, we
   see that the acceleration of the wall is very small compared to this UV cutoff in the field theory,
   ${ H \over \Lambda_{UV} } \sim { R_{AdS_4} \over R_{dS_3}} \sim e^{-\rho_0}$.

  The situation is very similar if we consider a thick brane. Let us first consider a thick brane
  whose tension is close to the critical value so that again we have a large range of $\rho$ where the
  geometry is well approximated by $AdS$. The scalar field   tries to approach the minimum of
  the potential but it is still  non-zero at $\rho=0$  \CdeL.
  In the dual field theory description, the fact that we have a scalar field which is approaching the
   minimum corresponds to the fact that we have a CFT deformed by an irrelevant operator. The effects of
   this irrelevant operator become of order one  where the thick  brane is localized. This is the
    UV cutoff region for the dual field theory.
     The effects of this irrelevant deformation become smaller as we go to the
   IR, but when we reach the $dS_3$ Hubble scale, this
   irrelevant deformation has not yet died completely. This scale corresponds to $\rho \sim 0$ in the
   above geometry.

\ifig\Expanding{ In (a) we have the Euclidean solution, with $\rho=0$ marking the center of the $AdS$ region.  In (b) we see the Lorentzian solution in the
thin wall regime. We can continue past $\rho =0$ into an FRW cosmology with $H_3$ slices. This FRW
cosmology expands and then contracts again. In the thin wall approximation the space is $AdS$ and
the collapse is a coordinate singularity.
 In (c) we see the Lorentzian solution in the thick wall regime.
 The nonzero energy density at $\rho =0$ or $\tau =0$, which is due to the displacement of
  the field from the minimum of the potential,
  generically produces a singularity at $\tau = \pi$.
 (d) Same decay described by an Euclidean solution with an   end of the world brane.
The false vacuum is outside and there is nothing inside.  (e) End of world brane picture for a Lorentzian solution.  In this case we have ``nothing'' inside the wall, but we have a field
 theory on the wall. Here we drew the picture for a flat space false vacuum, but similar pictures can
 be drawn for $AdS$ or $dS $ false vacua.
  } {\epsfxsize3.0in\epsfbox{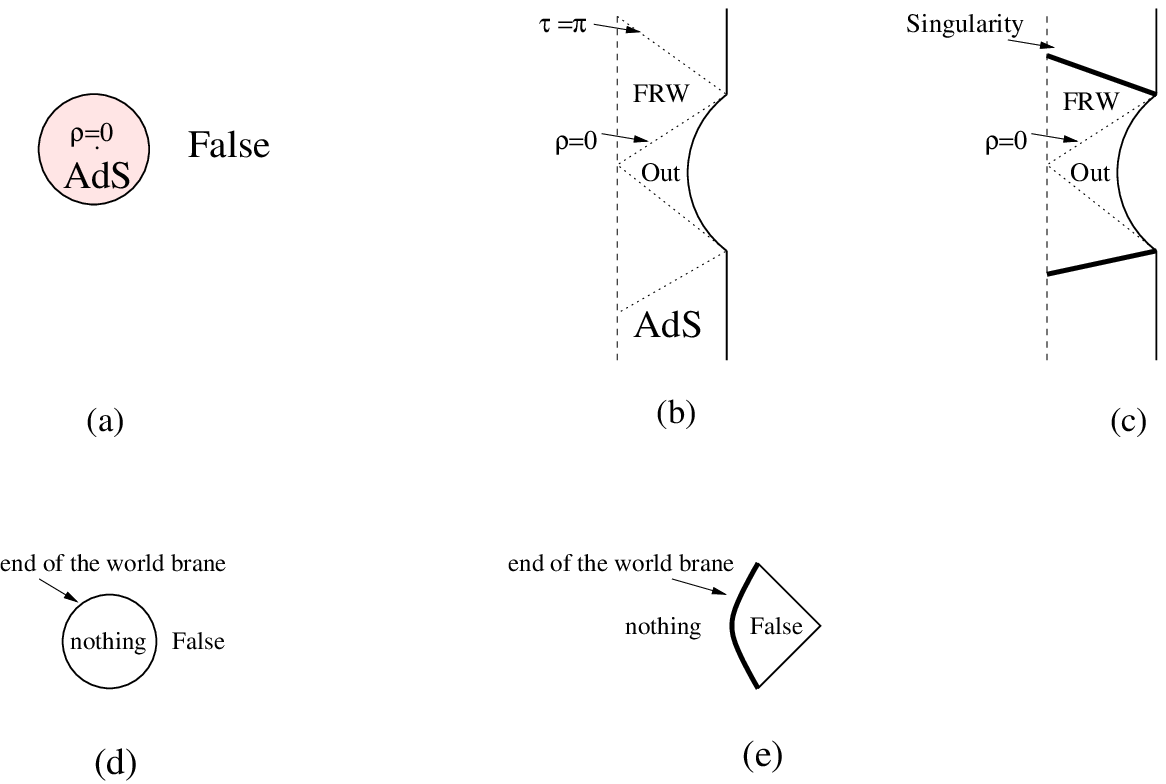}}

  As shown in \CdeL ,  we can continue behind the region $\rho =0$ into an FRW
    geometry which,
   as a first approximation, is
   \eqn\nege{
   ds^2 = R^2_{AdS} [ - d\tau^2 + \sin^2 \tau  ds^2_{H_3}]
  }
    where $\tau =0$ is the same as $\rho=0$. This corresponds to $AdS_4$ in hyperbolic slices. The universe
    expands and collapses again at $\tau = \pi$. If we were in $AdS$ this would be just a coordinate
    singularity. However, as  pointed out in \refs{\CdeL,\AbbottKR}, the fact that the scalar field is slightly displaced
    from the minimum at $\tau =0$  generically
leads to a singularity as we approach $\tau = \pi$. See   \Expanding .

The CFT gives a non-singular description of the decay process.
 In other words, if we had a CFT with
a known $AdS$ dual and we now put this CFT on de-Sitter space and we add an irrelevant deformation, then
we will also find a similar singular space. However, in this case the CFT description
is perfectly non-singular\foot{A field theory with an irrelevant deformation is not well defined at all energy scales. However, here we are only interested in the low energy behavior of such field theory. }.
 We do not expect singularities for field theories in de-Sitter space\foot{
For a dissident view see
 \Polyakov .}.   The gauge
 gravity duality on de-Sitter space was studied in \lref\BuchelWF{
  A.~Buchel,
  %``Gauge / gravity correspondence in accelerating universe,''
  Phys.\ Rev.\  D {\bf 65}, 125015 (2002)
  [arXiv:hep-th/0203041].
  %%CITATION = PHRVA,D65,125015;%%
}
%\AharonyCX
\lref\AharonyCX{
  O.~Aharony, M.~Fabinger, G.~T.~Horowitz and E.~Silverstein,
  %``Clean time-dependent string backgrounds from bubble baths,''
  JHEP {\bf 0207}, 007 (2002)
  [arXiv:hep-th/0204158].
  %%CITATION = JHEPA,0207,007;%%
}
%\BalasubramanianAM
\lref\BalasubramanianAM{
  V.~Balasubramanian and S.~F.~Ross,
  %``The dual of nothing,''
  Phys.\ Rev.\  D {\bf 66}, 086002 (2002)
  [arXiv:hep-th/0205290].
  %%CITATION = PHRVA,D66,086002;%%
}
%\RossCB
\lref\RossCB{
  S.~F.~Ross and G.~Titchener,
  %``Time-dependent spacetimes in AdS/CFT: Bubble and black hole,''
  JHEP {\bf 0502}, 021 (2005)
  [arXiv:hep-th/0411128].
  %%CITATION = JHEPA,0502,021;%%
}
%\CaiMR
\lref\CaiMR{
  R.~G.~Cai,
  %``Constant curvature black hole and dual field theory,''
  Phys.\ Lett.\  B {\bf 544}, 176 (2002)
  [arXiv:hep-th/0206223].
  %%CITATION = PHLTA,B544,176;%%
}
%\BalasubramanianBG
\lref\BalasubramanianBG{
  V.~Balasubramanian, K.~Larjo and J.~Simon,
  %``Much ado about nothing,''
  Class.\ Quant.\ Grav.\  {\bf 22}, 4149 (2005)
  [arXiv:hep-th/0502111].
  %%CITATION = CQGRD,22,4149;%%
}
%\HeJI
\lref\HeJI{
  J.~He and M.~Rozali,
  %``On Bubbles of Nothing in AdS/CFT,''
  JHEP {\bf 0709}, 089 (2007)
  [arXiv:hep-th/0703220].
  %%CITATION = JHEPA,0709,089;%%
}
%\HutasoitXY
\lref\HutasoitXY{
  J.~A.~Hutasoit, S.~P.~Kumar and J.~Rafferty,
  %``Real time response on dS_3: the Topological AdS Black Hole and the
  %Bubble,''
  JHEP {\bf 0904}, 063 (2009)
  [arXiv:0902.1658 [hep-th]].
  %%CITATION = JHEPA,0904,063;%%
}
%\MarolfTG
\lref\MarolfTG{
  D.~Marolf, M.~Rangamani and M.~Van Raamsdonk,
  %``Holographic models of de Sitter QFTs,''
  arXiv:1007.3996 [hep-th].
  %%CITATION = ARXIV:1007.3996;%%
}
 \refs{\HawkingDA,\BuchelTJ,\BuchelWF,\BuchelKJ,\AharonyCX,\BalasubramanianAM,\CaiMR,\RossCB,\AlishahihaMD,\BalasubramanianBG,\HirayamaJN,\BuchelEM,\HeJI,\HutasoitXY,\MarolfTG}, and
 references therein.
 Thus, the singularity is resolved in the same way that the black hole singularity is resolved by
 AdS/CFT. Namely, there is an ``in principle description'' via the field theory, but the concrete
 field theory computation  that describes a local observer near the singularity remains mysterious.

The question of whether we can continue time past the crunch is related to the question of whether we can continue
time past the time that the expanding brane hits the boundary of the false vacuum.
That is an interesting question, but it seems that to answer this question we need to go beyond the degrees
of freedom of the CFT, i.e. it involves energies above the UV cutoff of  the field theory.

Here we have discussed the case where the potential has an $AdS$ {\it minimum}. However, it can
 also have an $AdS$
 maximum,
as long as the curvature of the potential (or mass of the tachyon) obeys the Breitenlohner-Freedman bound
%\BreitenlohnerJF
\ref\BreitenlohnerJF{
  P.~Breitenlohner and D.~Z.~Freedman,
  %``Stability In Gauged Extended Supergravity,''
  Annals Phys.\  {\bf 144}, 249 (1982).
  %%CITATION = APNYA,144,249;%%
}. In
this case, the scalar field corresponds to a relevant deformation of the field theory and the scalar field
becomes {\it larger} as we approach $\rho =0$. In some cases, these relevant deformations can remove the
singularity completely by producing a mass gap which is above the $dS_3$ Hubble scale. In theories with
gravity duals, this often appears as a kind of end of the world brane that is cutting off the geometry at
 a finite warp factor of the $dS_3$ slices, see \refs{\HutasoitXY,\MarolfTG} for recent discussions.

\newsec{Conclusions}

This paper pointed out that vacuum decay processes where we produce $AdS$ regions can
conceptually be viewed as bubble of nothing decays,
 by replacing the whole $AdS$ region by an effective field theory
that lives on the domain wall. The particular CFT that we obtain depends on the theory we are considering.
Note that this field theory is defined with a  UV cutoff. The physics at the UV cutoff is encoded by the precise
bulk brane solution which joins the $AdS$ region into the false vacuum.
In cases where there is a large hierarchy between $R_{AdS}$ and $R_{dS}$ the field theory modes can
be   sharply separated from the dynamics of the thick brane. In cases where there is no sharp separation, the acceleration of the brane is comparable to the UV cutoff of the field theory. It is also likely that
  we can still view the setup as dual to an end of the world brane, but we do not have a clear argument.

The false vacuum itself
does not appear to be well defined, since it can decay!.
 Thus we do not expect to have the field theory defined with a
precision better than that of the vacuum where it lives. The false vacuum can be flat, AdS, or dS, but
we are dicussing   decays into $AdS$ spaces. Unstable $AdS$ vacua were studied from the point of view of AdS/CFT
in
%\BarbonGN
\ref\BarbonGN{
  J.~L.~F.~Barbon and E.~Rabinovici,
  %``Holography of AdS vacuum bubbles,''
  JHEP {\bf 1004}, 123 (2010)
  [arXiv:1003.4966 [hep-th]].
  %%CITATION = JHEPA,1004,123;%%
} and references therein.
 Note that the false vacuum will produce multiple bubbles that will collide. The description of these collisions
 takes us away from the field theory regime, and one would have to treat it using the full bulk
 solution. Thus, in our discussion we have ignored such collisions.
\lref\HertogHU{
  T.~Hertog and G.~T.~Horowitz,
  %``Holographic description of AdS cosmologies,''
  JHEP {\bf 0504}, 005 (2005)
  [arXiv:hep-th/0503071].
  %%CITATION = JHEPA,0504,005;%%
}
%\HorowitzWM
\lref\HorowitzWM{
  G.~Horowitz, A.~Lawrence and E.~Silverstein,
  %``Insightful D-branes,''
  JHEP {\bf 0907}, 057 (2009)
  [arXiv:0904.3922 [hep-th]].
  %%CITATION = JHEPA,0907,057;%%
}
%\HertogRZ
\lref\HertogRZ{
  T.~Hertog and G.~T.~Horowitz,
  %``Towards a big crunch dual,''
  JHEP {\bf 0407}, 073 (2004)
  [arXiv:hep-th/0406134].
  %%CITATION = JHEPA,0407,073;%%
  }
Finally, we should point out that field theories in de-Sitter with a small relevant deformations are an
interesting situation for understanding  big crunch singularities. Very closely related solutions
were studied in \refs{\HertogRZ,\HertogHU}\foot{We can reinterpret the O(1,3) symmetric solutions
in \refs{\HertogRZ,\HertogHU} as being dual to three dimensional field theories on $dS_3$ with a mass deformation. The boundary conditions are different from the ones in \refs{\HertogRZ,\HertogHU}, but the
actual solutions are the same. This is explained better in Appendix A}.
%Similar setups were considered in \HorowitzWM .  similar setups where studied in
%\HertogHU

 { \bf Acknowledgments }

We are   grateful to N. Arkani Hamed, T. Banks, G. Horowitz  and E. Silverstein  for discussions.

This work   was  supported in part by U.S.~Department of Energy
grant \#DE-FG02-90ER40542.

\appendix{A}{ Field theories in de-Sitter and cosmological  singularities \foot{This appendix
was added in January 2011, upon the suggestion of T. Banks.} }

\ifig\FieldTheory{ In (a) we see a Euclidean solution with $SO(4)$ symmetry corresponding
to a field theory on $S^3$ with a relevant deformation. In (b) we can see the scalar field
profile through a cross section of the solution indicated by the red line in (a).
In (c) we schematically depict the lorentzian continuation of the solution. $\rho =0$ becomes
a lightcone. Outside the lightcone we can foliate the spacetime with $dS_3$ slices on which the
$SO(1,3)$ symmetry acts. Inside the light cone $SO(1,3)$ acts on  $H_3$ slices. These slices
  expand and contract into
a singularity.
  } {\epsfxsize3.0in\epsfbox{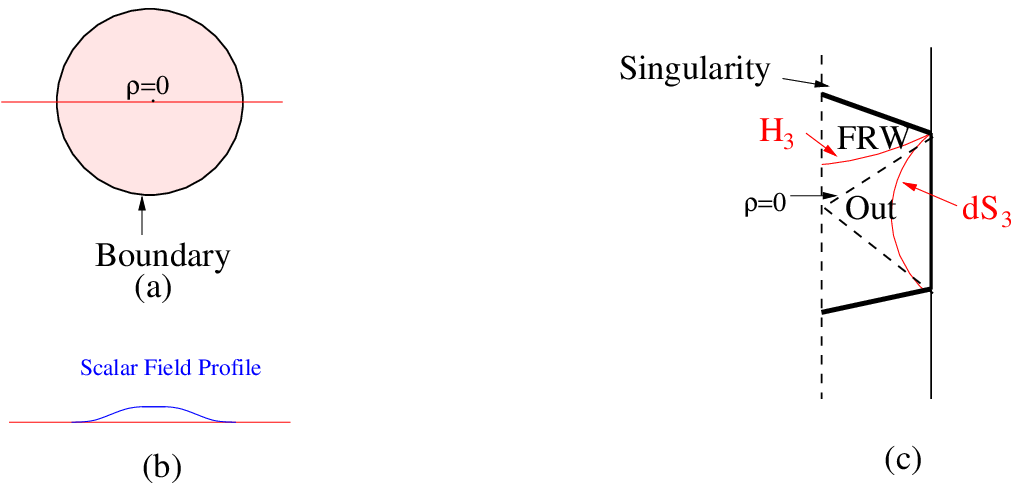}}

In the main part of the paper we discussed an
interpretation of Coleman de Luccia AdS bubbles that arise
through vacuum decay.
Here we give a short discussion  field theories in de-Sitter space and their gravity duals.
See \refs{\HawkingDA-\MarolfTG} for further discussion.  This is
a completely well defined setup,   which contains similar geometries in the interior.
We start with a conformal field theory in de-Sitter. We then add a
relevant deformation to the theory. For example, we can consider the theory describing M2 branes and
add a mass deformation, corresponding to a dimension one operator. If the mass term is small enough
the resulting field theory is   stable. By a small mass, we mean a   mass small compared
to the Hubble scale of the three dimensional de-Sitter space.
The gravity solution is
the $O(1,3)$ symmetric solution in  \refs{\HertogRZ,\HertogHU}. The geometry of these solutions
is summarized in \FieldTheory .  The Euclidean solution corresponds to the field
theory on $S^3$. For a small mass the geometry is a small deformation of Euclidean $AdS_4$ (or hyperbolic
space) with a scalar field that approaches zero at the boundary and it is non-zero, but relatively small,
everywhere else, see \FieldTheory (b).
The solutions in \refs{\HertogRZ,\HertogHU} contain a bulk scalar field, which, for large
$\rho$ behaves as
\eqn\behavl{
\phi(\rho) \sim  \alpha e^{- \rho } + \beta e^{-2 \rho }
}
with both $\alpha$ and $\beta $ nonzero. The quantization that makes this field dual to an operator
  of dimension one on the boundary corresponds to the one where we view $\beta$ as the parameter
we fix at the $AdS$ boundary. Thus $\alpha$ corresponds to a vev and $\beta $ to the boundary condition
%\KlebanovTB
\ref\KlebanovTB{
  I.~R.~Klebanov and E.~Witten,
  %``AdS/CFT correspondence and symmetry breaking,''
  Nucl.\ Phys.\  B {\bf 556}, 89 (1999)
  [arXiv:hep-th/9905104].
  %%CITATION = NUPHA,B556,89;%%
}.
This boundary condition is adding a term of the form $\beta {\cal O}$ to the field theory lagrangian, where
${\cal O}$ is an operator of dimension one.  We can embed these
solutions into eleven dimensions so that they are asymptotic to  $AdS_4 \times S^7/Z_k$
%\DuffGH
\lref\DuffGH{
  M.~J.~Duff and J.~T.~Liu,
  %``Anti-de Sitter black holes in gauged N = 8 supergravity,''
  Nucl.\ Phys.\  B {\bf 554}, 237 (1999)
  [arXiv:hep-th/9901149].
  %%CITATION = NUPHA,B554,237;%%
}
\refs{\DuffGH,\HertogHU}.
The the dual field theory is known %\AharonyUG
\ref\AharonyUG{
  O.~Aharony, O.~Bergman, D.~L.~Jafferis and J.~Maldacena,
  %``N=6 superconformal Chern-Simons-matter theories, M2-branes and their
  %gravity duals,''
  JHEP {\bf 0810}, 091 (2008)
  [arXiv:0806.1218 [hep-th]].
  %%CITATION = JHEPA,0810,091;%%
}, and the
operator has a form given by ${\cal O} = \alpha^I_{J}  Tr[ C^{I} (C^{J} )^\dagger ] $
where $\alpha^I_{J}$
are some constants, chosen so that the operator is real, $I=1,\cdots 4$ and $\alpha^I_I =0$.
For example, we can choose
$\alpha^1_{1} = - \alpha^2_2 =1$ and the rest zero. This operator gives mass squared terms
 which are positive for
some fields and negative for others. However, all fields have a positive contribution due to the
conformal coupling to the positive curvature of $S^3$ or $dS_3$. Thus, as long as the coefficient
of ${\cal O}$ is small enough, the theory is stable.
 The classical solutions are the same as those considered in
 \refs{\HertogRZ,\HertogHU} but the interpretation is different. The quantum corrections around
 the classical solutions would also be different since \refs{\HertogRZ,\HertogHU} were choosing
 different boundary conditions for the fields. With the boundary conditions in \refs{\HertogRZ,\HertogHU}
 the setup was unstable. With the  boundary conditions chosen here the theory is stable  for  small enough
 masses.

 When we continue this solution to Lorentzian signature the origin of the Euclidean
solution maps to a light cone. In the interior of the lightcone the $SO(1,3)$ symmetry acts on the spatial
slices which have an $H_3$ geometry. In the interior the solution is time dependent and gives rise to
a crunch singularity. The scalar field has a negative mass squared and a potential which is unbounded below.

In the main part of the paper we discussed potentials which had minima (rather than maxima) and
for that reason we considered irrelevant perturbations. In a field theory context we cannot consider
irrelevant perturbations at the UV boundary, since the field theory has to be well defined in the UV.
This is, of course, not a problem for the situation in the bulk of the paper where we had a UV cutoff. Of course, we can have an irrelevant perturbation if we UV complete the theory in a suitable way (via another UV CFT, for example).

 \listrefs

\bye